# The Architecture of Mr. DLib's Scientific Recommender-System API

Joeran Beel[1,2], Andrew Collins[1] and Akiko Aizawa[2]

[1] Trinity College Dublin, School of Computer Science & Statistics, ADAPT Centre, Ireland*
[2] National Institute of Informatics Tokyo, Digital Content and Media Sciences Division, Japan

`beelj@tcd.ie, ancollin@tcd.ie, aizawa@nii.ac.jp`

**Abstract.** Recommender systems in academia are not widely available. This may be in-part due to the difficulty and cost of developing and maintaining recommender systems. Many operators of academic products such as digital libraries and reference managers avoid this effort, although a recommender system could provide significant benefits to their users. In this paper, we introduce Mr. DLib's "Recommendations as-a-Service" (RaaS) API that allows operators of academic products to easily integrate a scientific recommender system into their products. Mr. DLib generates recommendations for research articles but in the future, recommendations may include call for papers, grants, etc. Operators of academic products can request recommendations from Mr. DLib and display these recommendations to their users. Mr. DLib can be integrated in just a few hours or days; creating an equivalent recommender system from scratch would require several months for an academic operator. Mr. DLib has been used by GESIS' Sowiport and by the reference manager JabRef. Mr. DLib is open source and its goal is to facilitate the application of, and research on, scientific recommender systems. In this paper, we present the motivation for Mr. DLib, the architecture and details about the effectiveness. Mr. DLib has delivered 94m recommendations over a span of two years with an average click-through rate of 0.12%.

## 1 Introduction

Scientific recommender systems automate information filtering in academia. They can therefore help to decrease information overload in academia. We define a 'scientific recommender system' as a software system that identifies a scientist's information need and recommends entities that satisfy that information need. Recommendable entities include research-articles [1–3], citations [4,5], call for papers [6], journals [7], reviewers [8], potential collaborators [9–11], genes and proteins [12], and research projects [13].

* This publication emanated from research conducted with the financial support of Science Foundation Ireland (SFI) under Grant Number 13/RC/2106.

The full potential of recommender systems in academia is not yet developed because not every scientist uses or has access to a scientific recommender system. Only a few reference managers such as Mendeley [3,14,15], Docear [16–18], and ReadCube[1] have integrated recommender systems, as have some scholarly search engines and digital libraries such as Google Scholar [19] and PubMed [20]. However, many services in academia (reference managers etc.) do not yet offer recommender systems. Consequently, users of such services still face the problem of information overload. We assume that most academic operators do not have the resources or skills to develop and maintain a recommender system.

We introduced "Mr. DLib", a scientific recommender-system as-a-service, previously [21,22]. Mr. DLib was originally developed as a **M**achine-**r**eadable **D**igital **Lib**rary at the University of California, Berkeley, and introduced in 2011 at the Joint Conference of Digital Libraries [22]. The original goal of Mr. DLib was to provide access to scientific literature in a machine-readable format. However, we decided to focus the future development more on related-article recommendations as-a-service (RaaS). The RaaS enables operators of, for example, reference managers or digital libraries to easily integrate a recommender system into their existing product. The operators do not need to develop and maintain a recommender system themselves.

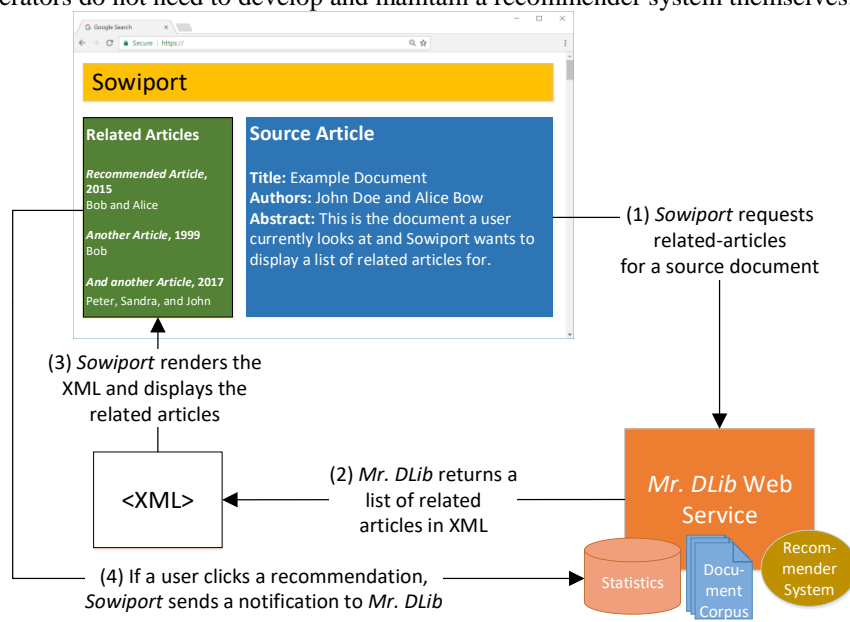

**Fig. 1.** Illustration of Mr. DLib's recommendation process

Fig. 1 illustrates the recommendation process. (1) A partner of Mr. DLib – in this case the academic search engine GESIS' Sowiport [23,24] – requests a list of related articles for an input document that is currently browsed by a user on Sowiport's search engine. (2) Mr. DLib generates a list of related articles and returns the article's metadata in XML format. (3) The partner displays the related articles on its own website.

---
[1] http://blog.readcube.com/post/94059448547/feature-of-the-week-recommendations

By offering scientific-recommendations as-a-service, Mr. DLib helps to reduce information overload in academia in two ways:

1. Mr. DLib enables operators of academic services to easily integrate recommender systems within their products. A partner can integrate Mr. DLib within a few hours, whereas it would take several months to develop their own equivalent recommender system. Expert knowledge of recommender systems is not required to integrate Mr. DLib. This way, more operators of academic services can offer recommender systems to their users.
2. Mr. DLib is open to recommender system researchers [25]. They can, for example, test their recommendation algorithms through Mr. DLib. Mr. DLib also publishes its data [26,27]. Hence, Mr. DLib supports the community to develop more effective scientific recommender systems in general.

In this paper, we present Mr. DLib and its architecture in detail, compared to the previous publication, which was only a 2-page poster [21]. Presenting Mr. DLib and its architecture in detail will help researchers to better understand how and why we conduct our research about related-article recommender systems [28–32]; explain how the system works to organizations that are interested in using Mr. DLib; and help organizations who want to build their own recommender system.

## 2    System Overview & Stakeholders

Mr. DLib has five stakeholders and the following general functionality (**Fig. 2**)[2]:

1. *Content Partners* submit content that is recommended by Mr. DLib's recommender system. For instance, publishers may submit their publications, academic social networks their user profiles, and conference organizers their call for papers.
2. *Service Partners* receive recommendations from Mr. DLib to display to their users. The recommendations are generated on the servers of Mr. DLib. The service partner requests recommendations for a specific user via HTTP request through a Restful API. Mr. DLib then returns a machine-readable XML file that contains a list of recommendations that the partner processes and displays to users. Alternatively, we also provide a JavaScript client which partners can add to their website. This client automatically requests and displays recommendations.
3. *Users* receive recommendations through service partners' products.
4. *Research partners* may analyze the data of Mr. DLib. They may also use Mr. DLib as a 'living lab', allowing them to evaluate their novel recommendation approaches through Mr. DLib. Their recommendation approaches are used to generate recommendations for our service partners' users.

---

[2] Content and distribution partners may also be the same organization, for instance, when a digital library provides content that shall be recommended on their own website.

5. The *operators* of Mr. DLib – i.e. us. We build and maintain Mr. DLib. We also act as research partners; our main motivation is to conduct research in the field of scientific recommender systems.

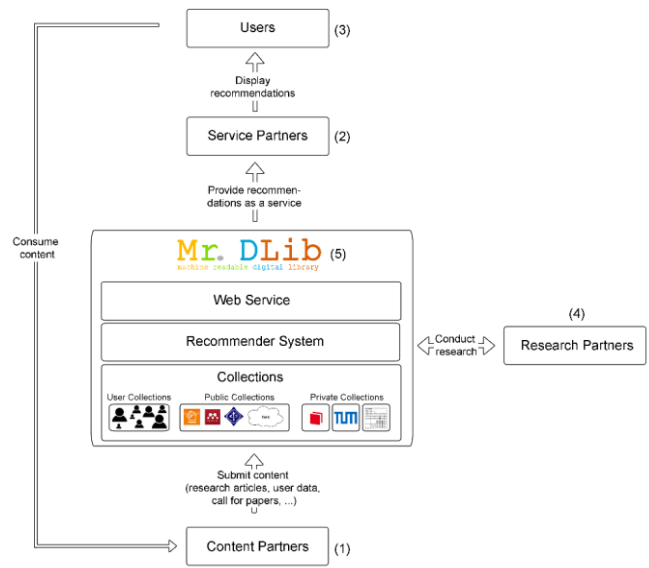

**Fig. 2.** Mr. DLib's Stakeholders and general System Overview

The partners' content is stored in collections, of which there are three types:

1. *Public collections* contain content that may be recommended to any service partner. Currently, Mr. DLib has one public collection from the CORE project [33–35]. This collection contains around 20 million documents[3] from three thousand research paper repositories[4].
2. *Private collections* are for content that is supposed to be recommended only to selected service partners. For instance, a university library might have little interest in distributing, or no rights to distribute, their content via third parties. With a private collection, only this library's users would receive recommendations for this content. Currently, Mr. DLib has one private collection, from the service and content partner Sowiport.
3. *User collections* store data of the partners' users. For instance, a reference manager might store its user data in such a collection to enable Mr. DLib calculating user-specific recommendations. Currently, Mr. DLib has not yet any partner that submits such data.

---

[3] The CORE corpus increases in size yearly; 90M document abstracts are available since our last content update and these documents will be recommended in the near future.
[4] https://core.ac.uk/repositories

## 3 Pilot Partners

### 3.1 GESIS' Sowiport

GESIS – Leibniz-Institute for the Social Sciences is the largest infrastructure institution for the Social Sciences in Germany. It is operating the portal Sowiport that pools and links social-science information from domestic and international providers, making it available in one place [23,24,36–38]. Sowiport currently contains 9.5 million references on publications and research projects. The documents in Sowiport comprise bibliographic metadata (such as authors, publishers, keywords), citation and reference information and roughly 1.3 million full text links. For each of the 9.5 million articles in Sowiport, a detail page exists. On each of these pages, recommendations are displayed from Mr. DLib (**Fig. 3**).

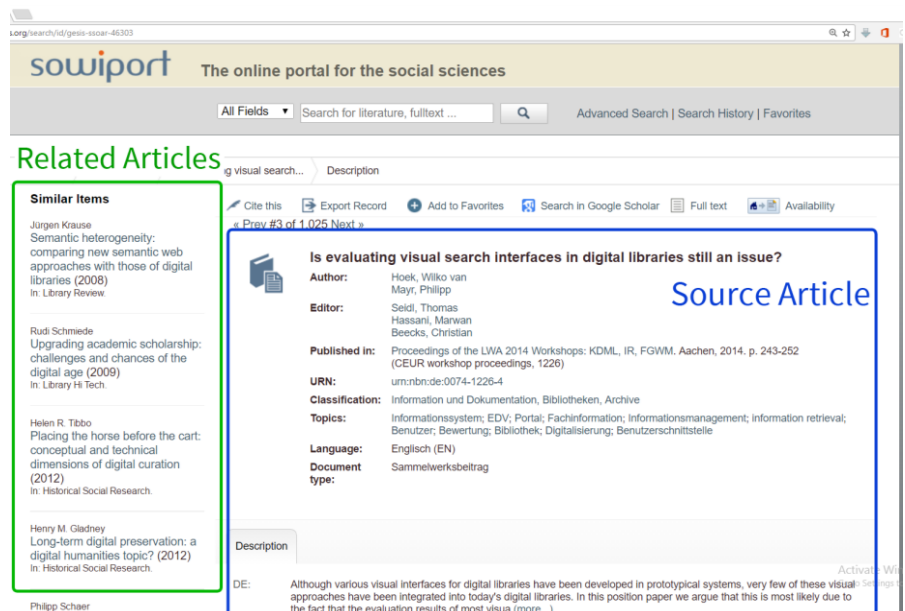

**Fig. 3.** Mr. DLib's "related articles" recommendations on GESIS' Sowiport

### 3.2 JabRef

JabRef is one of the most popular reference managers with millions of downloads over the past decade and tens of thousands active users [39][5]. The main interface of JabRef consists of a list with all articles a user intends to reference. A double click on an entry opens the editor window. In this editor window, users may select a "Related Articles" tab (**Fig. 4**). When this tab is selected, JabRef sends a request to Mr. DLib containing the document's title. If Mr. DLib has the input document in its database, Mr. DLib returns a list of related articles. If the document is not in Mr. DLib's database, the recommender system interprets the title as search query for Lucene and returns Lucene's search results as related articles.

---
[5] https://sourceforge.net/projects/jabref/files/jabref/stats/timeline?dates=2003-10-12+to+2016-08-16

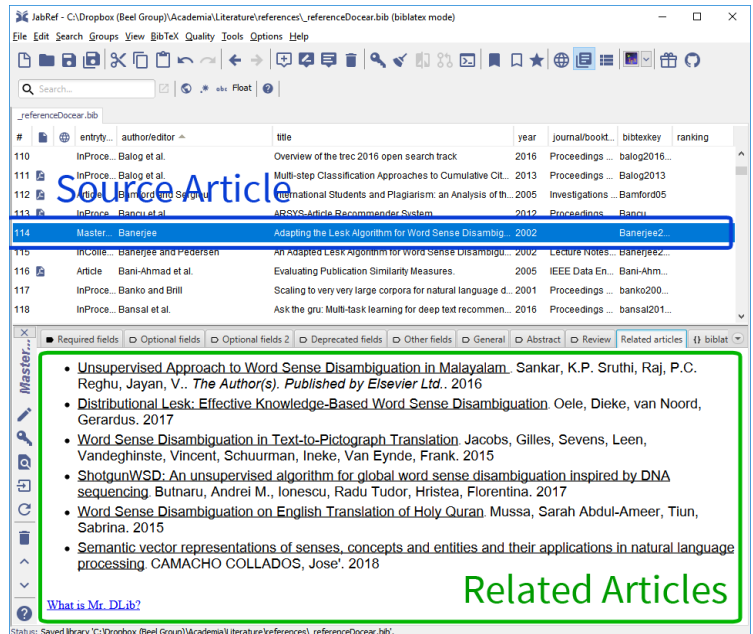

**Fig. 4.** Mr. DLib's "related articles" recommendations in JabRef

## 4  The Architecture in Detail

A high-level view of Mr. DLib's architecture is shown in **Fig. 6**. Here we describe each component of this architecture in detail.

Mr. DLib runs on two servers: one development system and one production system. Both are dedicated servers with almost identical specifications. They both have an Intel Core i7-4790K, 32 GB RAM, and 1TB SSD. The development system – on which resource-intensive tasks are performed such as parsing XML files and calculating document embeddings – has an added 2TB SATA.

Parsing all XML files of GESIS (60GB in size, containing 10 million documents) storing the relevant information in the database, and indexing the data in Lucene requires several weeks.

The Production system's specification allows Mr. DLib to be responsive to requests. 65% of recommendation requests are received, processed and responded to in less than 150ms, and 84% in less than 250ms (**Fig. 5**).

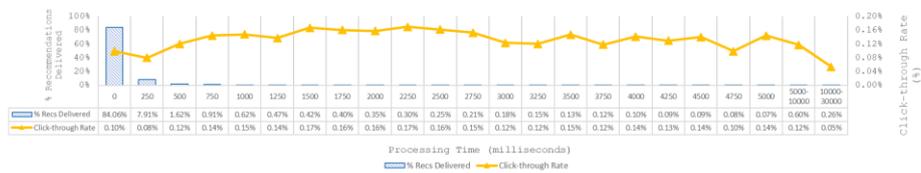

**Fig. 5.** The number of recommendations delivered, organized by their processing time (milliseconds). 84% of recommendations are processed within 250ms

The central element of Mr. DLib is its Master Data storage, namely a MySQL database. This database contains all relevant data including documents' metadata and statistics of delivered recommendations.

Our "Content Acquisition" process downloads partners' content once a month. Currently, Mr. DLib has only one partner with a private collection; GESIS provides their corpus of 9.5 million documents as a Solr XML export. The XML files are backed up on Mr. DLib's server and then the relevant metadata of the documents is stored in the database. Although, GESIS provides full-texts for some documents, Mr. DLib currently does not utilize it for recommendations due to storage and CPU constraints. In future we will use full-texts of documents for calculating recommendations, and for in-text citation analysis to calculate document similarities based on metrics such as CPA [40].

The CORE project's public collection increases in size frequently, and we periodically update our storage of its metadata.

Mr. DLib uses several recommendation frameworks to generate recommendations. We primarily use Apache Solr/Lucene for its fast search-response times, and for its "More like this" class. This class calculates content-based document similarities using TF-IDF. It also offers a configurable query parser. As well as Apache Lucene, we also use Gensim[6] to generate document embedding-based recommendations [41]. We plan to introduce more recommendation frameworks, namely Apache Mahout and LensKit. Every recommendation framework we use can retrieve required data from the Master Data storage.

Mr. DLib harnesses different recommendation approaches. As well as TF-IDF and document-embeddings, we also generate keyphrases for all articles in the corpus and make recommendations based on them. We further utilise stereotype, and most-popular recommendation algorithms. Our stereotype approach assumes the persona of a typical academic user and recommends documents suitable for that persona. Our most-popular approach recommends the most-popular documents from Sowiport. "Popularity" is measured by "Views", i.e. the most viewed articles on Sowiport's website, and by "Exports", i.e. the most exported documents on Sowiport's website.

Our TF-IDF, document-embedding, and key-phrase-based recommenders can use documents' titles and abstracts to find related documents for a given input document.

Mr. DLib also uses external recommendation APIs such as the *CORE Recommendation API* [42,43] to make recommendations.

---

[6] https://radimrehurek.com/gensim/

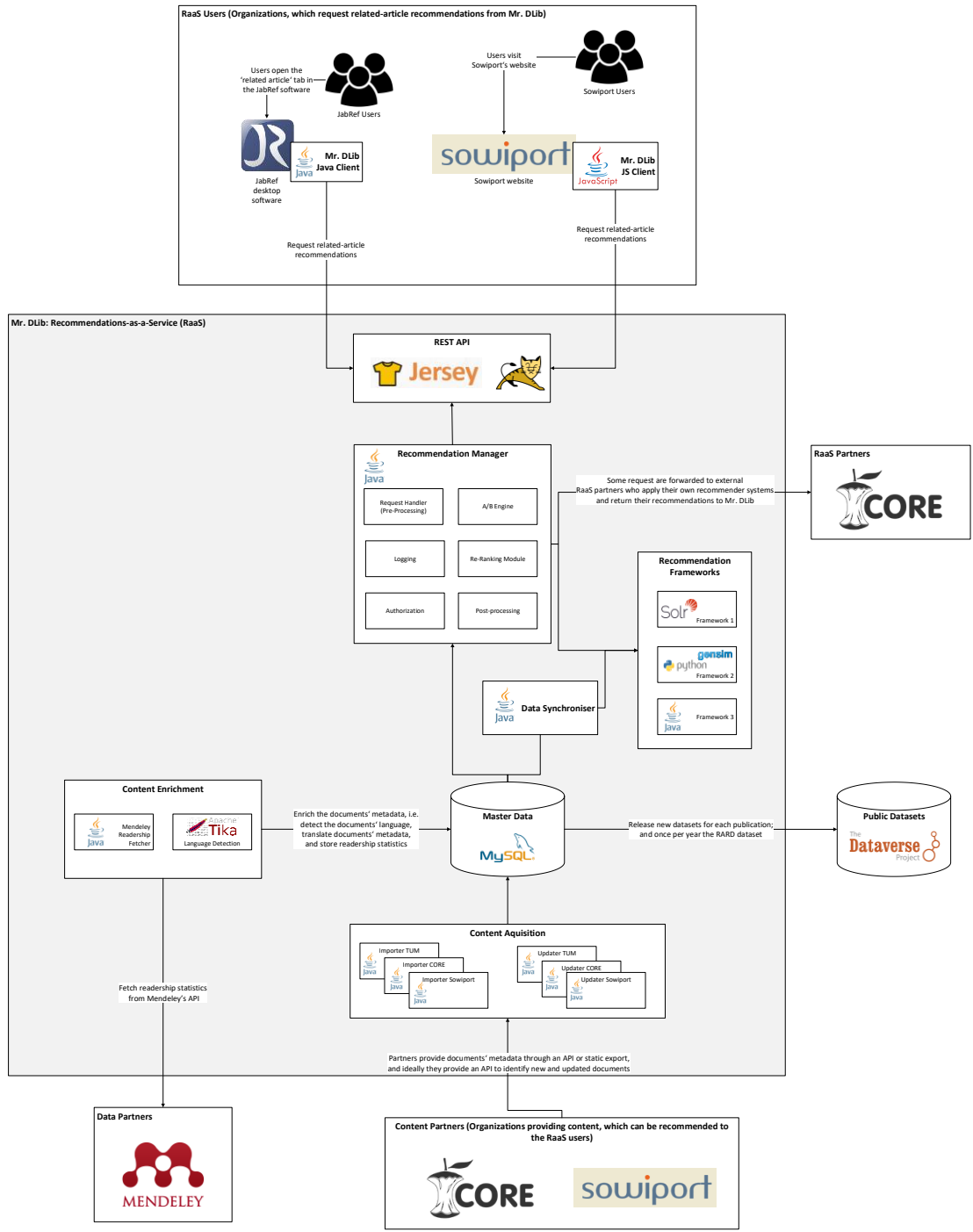

**Fig. 6.** Mr. DLib's architecture

Mr. DLib offers a RESTful API. A partner interacts with Mr. DLib via HTTP requests (typically GET requests). To retrieve recommendations, the partner calls https://api.mr-dlib.org/v1/documents/<partner-document_id>/related_documents/ and retrieves an XML response containing a list of related documents (**Fig. 7**). Mr. DLib's web service is realized with Apache Tomcat and JAVA Jersey. The proprietary "API Manager" writes some statistics to the database and forward the requests to the proprietary "Recommendation Manager".

**Fig. 7.** XML response containing a list of related documents

Our "Content Enrichment" process gathers data from external sources to enhance the recommendation process. For example, for each document we request readership statistics from Mendeley's API [44][7]. We can then optionally use readership statistics to re-rank recommendations based on the document's attributes on Mendeley. We further use Apache Tika's language detector to corroborate any language metadata in the corpuses.

The "Recommendation Manager" (JAVA) handles all processes related to recommendations. It looks up required data from the database (e.g. matches the partner's document id from the URL with Mr. DLib's internal document ID), decides which recommendation framework to use, which recommendation approach to use, calculates and stores statistics, and re-ranks recommendation candidates based on scientometrics or based on our experimental requirements.

Parameterization of all algorithms is managed by Mr. DLib's A/B testing engine. To take one example of a recommendation instance: The A/B engine may choose Apache Lucene and content-based filtering as a recommendation approach. It randomly selects whether to use 'normal keywords' or 'key-phrases' [45]. For each option, additional parameters are randomly chosen; e.g., when key-phrases are chosen, the engine randomly selects whether to use key-phrases from the 'title' or 'abstract'. Subsequently, the system randomly selects whether unigram, bigram, or trigram key-phrases are used. The system randomly selects how many key-phrases to use when calculating document relatedness. The A/B engine also randomly chooses which query parser to use (standardQP or edismaxQP). Finally, the engine selects whether to re-rank recommendations with readership data from Mendeley, and how many recommendations to return. All of these details are logged by Mr. DLib.

---
[7] http://dev.mendeley.com/

We want to ensure that we deliver good recommendations. Therefore, our A/B engine makes its 'random' choices with unequal probabilities. We do not want to deliver recommendations using an experimental algorithm with the same probability as our most effective algorithm, for example. Our probabilities are in-part defined according to our previous evaluations [28]. Approximately 90% of recommendations are delivered using our strongest algorithms, and 10% is allocated to various experimental algorithms and baselines.

In order to support the recommender system community, we periodically publish Mr. DLib recommendation log data. We have published two iterations of the Related-Article Recommendation Dataset (RARD)[8]. We released RARD I, which comprised 57.4 million recommendations, in 2017 [26]. We subsequently released RARD II in 2018 [27]; this iteration contains 64% more recommendations than RARD I, as well as 187% more features, 50% more clicks, and 140% more documents. The RARD datasets are unique in the scale and variety of recommender system meta-data that they provide. They allow researchers to benchmark their recommendation techniques, and to evaluate new approaches.

Mr. DLib is mostly developed in JAVA and uses standard tools and libraries whenever possible.

Mr. DLib's source code is published open source on GitHub under GPL2+ and Apache 2 license[9]. There is a public WIKI and volunteers are welcome to join the development. In the future, some code may be kept private or published under different licenses if data privacy or copyright of a partner requires this. This could be the case if, for instance, a crawler for a partner's data would reveal information about the partner, or their data, that the partner does not want to be public. Similarly, user specific data and partner content is not publicly available to ensure data privacy of users and copyrights of content partners.

The uptime of our development and production systems is monitored constantly using a third-party service[10]. We have had no significant outages since Mr. DLib's inception and work to maintain 100% uptime for our partner-facing production system.

---

[8] https://dataverse.harvard.edu/dataverse/Mr_DLib
[9] https://github.com/BeelGroup
[10] https://uptimerobot.com/

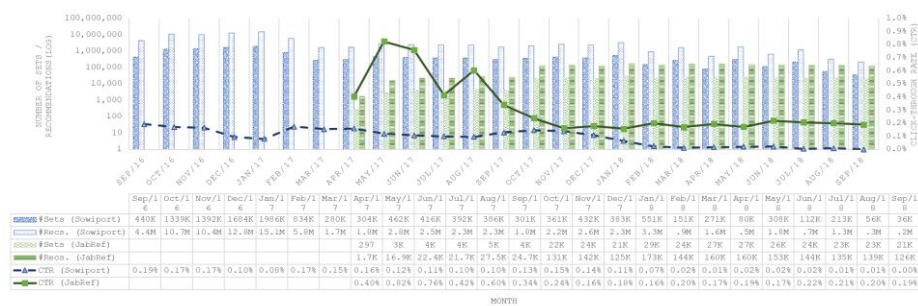

**Fig. 8.** The number of recommendations delivered, and click-through rates, for our two partners between September 2016 and September 2018

## 5 Usage Statistics

Between September 2016 and September 2018, Mr. DLib has delivered 94m recommendations to partners. Users clicked upon recommendations 113,954 times. This gives an overall-average click-through rate (CTR)[11] of 0.12%.

**Fig. 8** illustrates usage and user engagement for Sowiport and Jabref within this time period.

Our highest priority is to provide the best recommendations possible for our partners and for end-users, and to increase recommendation effectiveness. To this end, we have conducted many experiments which aim to examine recommendation effectiveness or to improve it. These experiments include: increasing recommendation-ranking accuracy based on Mendeley Readership data [44]; assessing the effect of position bias on user engagement [32]; assessing choice overload with respect to recommendation-set size [30]; evaluating stereotype and most-popular recommendation algorithms [28]; coordinating with research-partners to evaluate their own recommender system using Mr. DLib as a living-lab [25].

We keep extensive records of recommendation effectiveness by partner, algorithm, week, month, and so on. **Fig. 9** illustrates the overall effectiveness of our key classes of algorithm per month, between September 2016 and September 2018.

## 6 Related Work

In Academia, RaaS for related research articles are offered by a few organizations. BibTip [12] [46,47] and ExLibris bX [13] offer literature recommendations for digital libraries and both apply the same recommendation concept, namely co-occurrence-based recommendations [48]. BibTip and bX are for-profit companies that do not publish their recommender systems' source code, nor publish research results of their systems. In addition, both BibTip and bX only address digital libraries but no other academic service operators such as reference managers. A service similar to Mr. DLib was TheAdvisor [49], a citation recommender system that offered an API. However, the website has been defunct for several years[14]. The two most similar works to Mr. DLib are Babel[15] [50] and the CORE recommender[16] [34,43]. Babel is developed by researchers at DataLab, which is part of the Information School at the University of Washington. CORE is mostly developed by the Knowledge Media institute at The Open University. Both Babel and CORE are similar to Mr. DLib in many aspects: the motivation for the service, the architecture, the philosophy (open source), and the audience are similar to Mr. DLib. However, as far as we know neither of these services, for instance, has a living lab or publishes their data.

---

[11] We use click-through rate as a metric to gauge recommender effectiveness. This is the ratio of recommendations clicked, to recommendations delivered.
[12] http://www.bibtip.com/en
[13] http://www.exlibrisgroup.com/category/bXRecommender

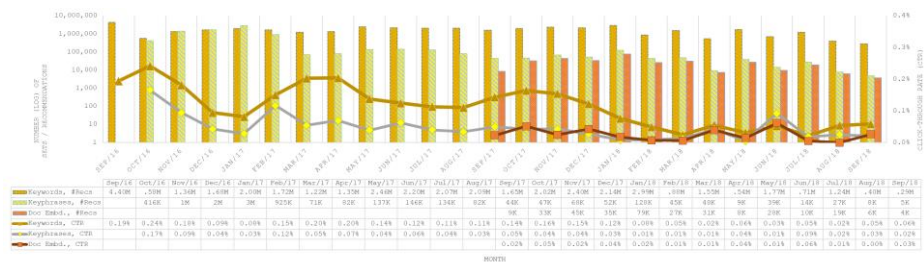

**Fig. 9.** Performances of our key classes of recommendation algorithm for each month between September 2016 and September 2018

## 7  Summary and Future Work

Many further developments are planned for Mr. DLib:

— Currently, Mr. DLib is recommending only research articles. In the future, Mr. DLib will also recommend other entities such as conference call for papers, journals, datasets, persons (experts, and potential collaborators), projects, and maybe also Wikipedia pages, academic news, blogs, presentations, and mathematical formulas.
— We want to have several distribution partners in each of the following categories: digital libraries, publishers, search engines, and reference managers. This will allow us to evaluate the effectiveness of recommendation approaches in diverse scenarios.
— Currently, Mr. DLib applies several content-based-filtering algorithms (terms, keyphrases, document embeddings, stereotype, most popular). In the future, we want to introduce collaborative filtering approaches. We further plan to introduce meta-learning-based approaches for algorithm selection [31,51], and ensemble-based approaches for algorithm weighting, to maximize recommendation effectiveness.

---

[14] http://theadvisor.osu.edu/

[15] http://babel.eigenfactor.org/

[16] https://core.ac.uk/

In addition, organizational improvements will be made. The website http://mr-dlib.org will be extended, to make it easier for external developers to contribute to the project, and more information for potential content and distribution partners must be provided. In the long-run some administration interface for the partners might be desirable.